\documentclass[twoside,twocolumn,9pt]{article}
\usepackage{extsizes}
\usepackage[super,sort&compress,comma]{natbib} 
\usepackage[version=3]{mhchem}
\usepackage[left=1.5cm, right=1.5cm, top=1.785cm, bottom=2.0cm]{geometry}
\usepackage{balance}
\usepackage{mathptmx}
\usepackage{sectsty}
\usepackage{graphicx} 
\usepackage{lastpage}
\usepackage[format=plain,justification=justified,singlelinecheck=false,font={stretch=1.125,small,sf},labelfont=bf,labelsep=space]{caption}
\usepackage{float}
\usepackage{fancyhdr}
\usepackage{fnpos}
\usepackage{amssymb}
\usepackage[english]{babel}
\addto{\captionsenglish}{%
  \renewcommand{\refname}{Notes and references}
}
\usepackage{array}
\usepackage{droidsans}
\usepackage{charter}
\usepackage[T1]{fontenc}
\usepackage[usenames,dvipsnames]{xcolor}
\usepackage{setspace}
\usepackage[compact]{titlesec}
\usepackage{hyperref}
\newcommand{\angstrom}{\mbox{\normalfont\AA}} 

\usepackage{epstopdf}

\definecolor{cream}{RGB}{222,217,201}

\begin{document}

\pagestyle{fancy}
\thispagestyle{plain}
\fancypagestyle{plain}{
\renewcommand{\headrulewidth}{0pt}
}

\makeFNbottom
\makeatletter
\renewcommand\LARGE{\@setfontsize\LARGE{15pt}{17}}
\renewcommand\Large{\@setfontsize\Large{12pt}{14}}
\renewcommand\large{\@setfontsize\large{10pt}{12}}
\renewcommand\footnotesize{\@setfontsize\footnotesize{7pt}{10}}
\makeatother

\renewcommand{\thefootnote}{\fnsymbol{footnote}}
\renewcommand\footnoterule{\vspace*{1pt}%
\color{cream}\hrule width 3.5in height 0.4pt \color{black}\vspace*{5pt}} 
\setcounter{secnumdepth}{5}

\makeatletter 
\renewcommand\@biblabel[1]{#1}            
\renewcommand\@makefntext[1]%
{\noindent\makebox[0pt][r]{\@thefnmark\,}#1}
\makeatother 
\renewcommand{\figurename}{\small{Fig.}~}
\sectionfont{\sffamily\Large}
\subsectionfont{\normalsize}
\subsubsectionfont{\bf}
\setstretch{1.125} 
\setlength{\skip\footins}{0.8cm}
\setlength{\footnotesep}{0.25cm}
\setlength{\jot}{10pt}
\titlespacing*{\section}{0pt}{4pt}{4pt}
\titlespacing*{\subsection}{0pt}{15pt}{1pt}

\fancyfoot{}
\fancyfoot[LO,RE]{\vspace{-7.1pt}\includegraphics[height=9pt]{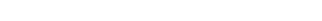}}
\fancyfoot[CO]{\vspace{-7.1pt}\hspace{13.2cm}\includegraphics{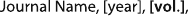}}
\fancyfoot[CE]{\vspace{-7.2pt}\hspace{-14.2cm}\includegraphics{head_foot/RF}}
\fancyfoot[RO]{\footnotesize{\sffamily{1--\pageref{LastPage} ~\textbar  \hspace{2pt}\thepage}}}
\fancyfoot[LE]{\footnotesize{\sffamily{\thepage~\textbar\hspace{3.45cm} 1--\pageref{LastPage}}}}
\fancyhead{}
\renewcommand{\headrulewidth}{0pt} 
\renewcommand{\footrulewidth}{0pt}
\setlength{\arrayrulewidth}{1pt}
\setlength{\columnsep}{6.5mm}
\setlength\bibsep{1pt}

\makeatletter 
\newlength{\figrulesep} 
\setlength{\figrulesep}{0.5\textfloatsep} 

\newcommand{\topfigrule}{\vspace*{-1pt}%
\noindent{\color{cream}\rule[-\figrulesep]{\columnwidth}{1.5pt}} }

\newcommand{\botfigrule}{\vspace*{-2pt}%
\noindent{\color{cream}\rule[\figrulesep]{\columnwidth}{1.5pt}} }

\newcommand{\dblfigrule}{\vspace*{-1pt}%
\noindent{\color{cream}\rule[-\figrulesep]{\textwidth}{1.5pt}} }

\makeatother

\twocolumn[
  \begin{@twocolumnfalse}
{\includegraphics[height=30pt]{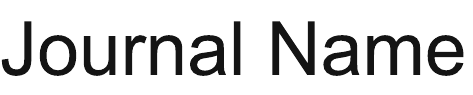}\hfill\raisebox{0pt}[0pt][0pt]{\includegraphics[height=55pt]{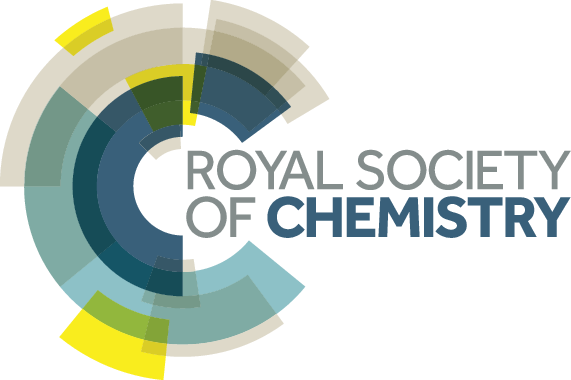}}\\[1ex]
\includegraphics[width=18.5cm]{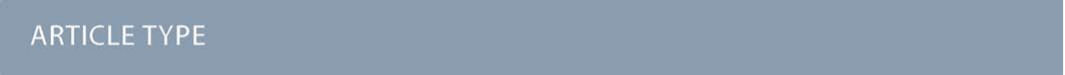}}\par
\vspace{1em}
\sffamily
\begin{tabular}{m{4.5cm} p{13.5cm} }

\includegraphics{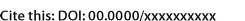} & \noindent\LARGE{\textbf{Interface energies of $\mathrm{Ga_2O_3}$ phases with the sapphire substrate and the phase-locked epitaxy of metastable structures explained $^\dag$}}\\
\vspace{0.3cm} & \vspace{0.3cm} \\

 & \noindent\large{Ilaria Bertoni$^a{\ast}$, Aldo Ugolotti$^a$, Emilio Scalise$^a$, Roberto Bergamaschini$^a$, and Leo Miglio$^a$} \\


\includegraphics{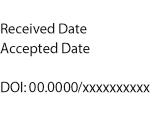} & \noindent\normalsize{Despite the extensive work carried out on the epitaxial growth of $\mathrm{Ga_2O_3}$, a fundamental understanding of the nucleation of its different metastable phases is still lacking. Here we address the role of interface energies by Density Functional Theory calculations of $\mathrm{\alpha}$, $\mathrm{\beta}$ and $\mathrm{\kappa-Ga_2O_3}$ on (0001) $\mathrm{Al_2O_3}$ substrates, and different $\mathrm{Ga_2O_3}$ interlayers. In conjunction to surface energies and misfit strain contribution, we demonstrate that $\mathrm{\alpha-Ga_2O_3}$ is the preferred phase in 2D islands, when the low growth temperatures and the high growth rates hinder 3D island nucleation. This quantitatively explains the phase-locking in Mist-CVD experiments. }\end{tabular}

 \end{@twocolumnfalse} \vspace{0.6cm}

  ]

\renewcommand*\rmdefault{bch}\normalfont\upshape
\rmfamily
\section*{}
\vspace{-1cm}


\footnotetext{\textit{$^{a}$~Department of Materials Science, University of Milano-Bicocca, via Cozzi 55, 20125 Milan (Italy)}}

\footnotetext{\dag~Supplementary Information available: details on oriented bulk cells used for calculations, additional images of interface slabs, details on parameters for the calculation of the formation energy of $\mathrm{Ga_2O_3}$ 2D islands, files of the optimized geometries. See DOI: 00.0000/00000000.}


\section*{Introduction}
The monoclinic phase of gallium oxide ($\mathrm{\beta}$-$\mathrm{Ga_2O_3}$) has positioned itself as strong contender in next-generation high-power electronics \cite{review1, review2, power1}, due to some superior properties compared to the current leader, silicon carbide (4H-SiC). These advantages include wider bandgap (4.9 eV), providing a higher breakdown voltage, and moderate epitaxial growth temperatures, 700-800 °C by Metal-Organic Chemical Vapor Deposition (MOCVD), potentially enabling integration into the Si technology. However, while $\mathrm{\beta}$-$\mathrm{Ga_2O_3}$ is the stable crystal phase, epitaxial growth can yield different competing phases. Among these, the $\mathrm{\alpha}$ phase (rhombohedral) and the $\mathrm{\kappa}$ phase (orthorhombic, previously addressed as  $\mathrm{\epsilon}$-hexagonal) are particularly attractive for power electronics\cite{Nishinaka-k-domains, Coral-k-real-struct}. Understanding and controlling the growth of different $\mathrm{Ga_2O_3}$ phases is crucial to unlock its full potential for high-power device fabrication.

c-sapphire emerges as the preferred substrate for growing $\mathrm{Ga_2O_3}$ phases via heteroepitaxy, both due to its structural coherence with $\mathrm{\alpha}$-$\mathrm{Ga_2O_3}$ and its convenient cost compared to other potential substrates. The $\mathrm{\alpha}$ phase grows exposing the same surface as that of the substrate, i.e. the (0001) plane. The monoclinic $\mathrm{\beta}$ phase and the orthorhombic $\mathrm{\kappa}$ phase grow in three-fold rotational domains \cite{Nishinaka-k-domains}, due to the peculiar arrangement in layers of the oxygen atoms along the ($\bar{2}$01) plane of the former and the (001) plane of the latter \cite{growth_dir1, growth_dir2, growth_dir3, growth_dir4, growth_dir5, growth_dir6}, which nearly matches in symmetry the triangular oxygen network of c-sapphire.

\begin{figure}[h]
\centering
  \includegraphics[height=14cm]{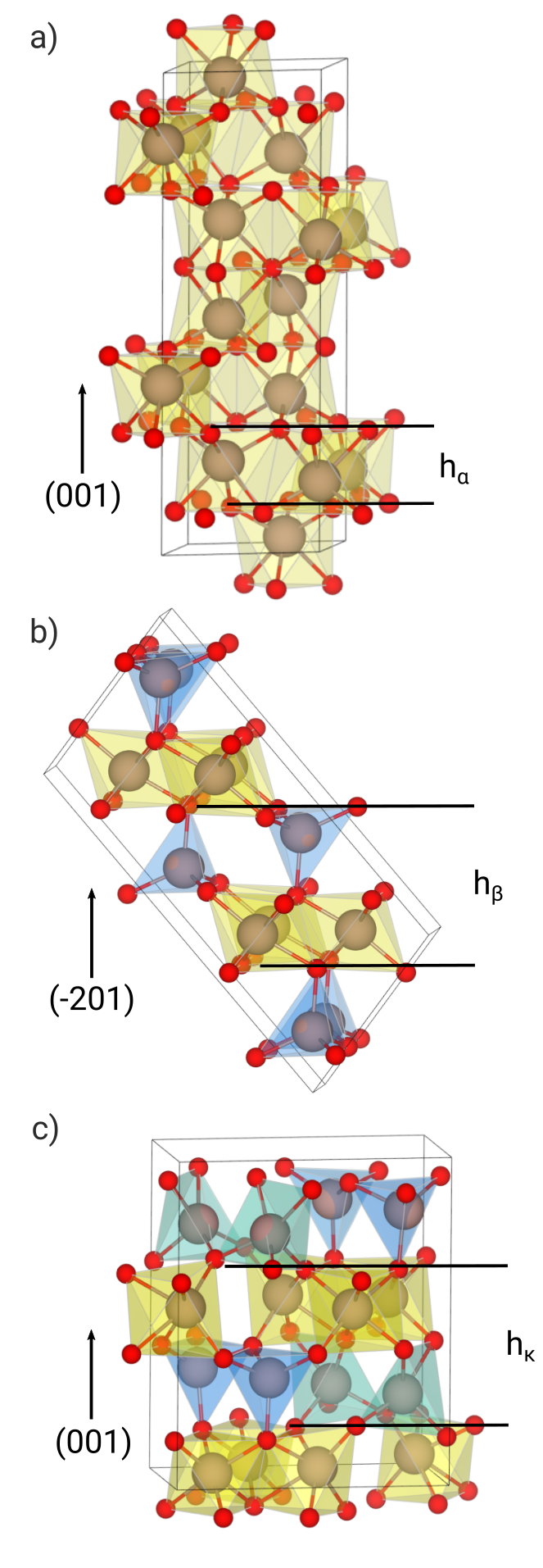}
  \caption{Optimized structures of the conventional cell of bulk $\mathrm{\alpha-}$ (a), $\mathrm{\beta-}$ (b) and $\mathrm{\kappa-}$(c) $\mathrm{Ga_2O_3}$. Ga and O atoms are shown with grey and red spheres, respectively. Tetrahedral, pentahedral and octahedral Ga-O coordination structures are shown through yellow, green and blue surfaces. The vertical direction of the image is aligned with growth direction of $\mathrm{Ga_2O_3}$ on c-sapphire substrate. For each phase the inequivalent epitaxial layer is indicated by $h$. }
  \label{fgr:bulk}
\end{figure}

As shown in Figure \ref{fgr:bulk}, the three phases are characterized by a different Ga–O coordination. While in $\mathrm{\alpha}$-$\mathrm{Ga_2O_3}$ all Ga atoms are coordinated with six O atoms forming octahedral cages, in $\mathrm{\beta}$ and $\mathrm{\kappa}$ phases 50\% and 25\% of Ga atoms are four-fold coordinated, respectively, forming tetrahedral complexes. It is worth pointing out that in the $\mathrm{\kappa}$ phase one third of Ga atoms in the octahedral cage may result five-fold coordinated, depending on the coordination cut-off radius, as one oxygen atom is 0.45  \angstrom  more distant than the others, due to the distortion of the octahedral cage (see the SI of our previous paper on $\mathrm{Ga_2O_3}$ surfaces, \cite{Bertoni}). The vertical axis in Figure \ref{fgr:bulk} indicates the actual growth direction and the alternate stacking of oxygen atoms layers (in red) is evident.
Moreover, all $\mathrm{Ga_2O_3}$ phases experience a degree of mismatch with the sapphire lattice. This mismatch ranges from about 4\% for $\mathrm{\alpha}$-$\mathrm{Ga_2O_3}$ to a larger and anisotropic mismatch for $\mathrm{\beta}$ and $\mathrm{\kappa}$ phases\cite{Bertoni}. The mismatch creates strain at the interface between the $\mathrm{Ga_2O_3}$ film and the sapphire substrate. Different mechanisms, such as misfit dislocations (especially for $\mathrm{\alpha}$-$\mathrm{Ga_2O_3}$) and defected boundaries between rotational domains (for $\mathrm{\beta}$ and $\mathrm{\kappa}$ phases), may be responsible for strain release, without considering elastic relaxation in three-dimensional islands  \cite{growth_dir1, Kaneko_2012}.  At the moment, very little is experimentally assessed in terms of such strain relief.

The role of the interface and its related kinetic issues appears to be evident by the fact that various growth methods (mist-CVD, MOCVD, High-Vacuum VPE, MBE, PLD, Halide-VPE) and diverse growth conditions \cite{growth1, growth2, growth3, growth4, Fornari2021} lead to different phases. In this complex picture, some trends are sufficiently clear: low temperatures (and/or high growth rates) favor the $\mathrm{\alpha}$ phase, intermediate temperature values may favour the $\mathrm{\kappa}$ phase, whereas conditions closer to equilibrium induce the stable $\mathrm{\beta}$ phase. In a recent review by Kaneko et al. \cite{growth4}, primarly focusing on the investigation of low temperature mist-CVD growth of $\mathrm{\alpha-Ga_2O_3}$, the authors suggest that the (meta) stability of this phase at low temperature is produced by a "phase locking" at the interface. Such a mechanism would kinetically favour the phases with a better structural match with the substrate, independently of other thermodynamic factors that come into play with temperature (and surface diffusion length).

However, no quantitative validation of this intuition was possible due to the lack of interface energies. Therefore, the ultimate goal of our work is to provide interface energies for a deeper understanding of the stabilization of the different $\mathrm{Ga_2O_3}$ phases on sapphire and $\mathrm{Ga_2O_3}$ interlayers, at the same time providing a solid framework for growth models to be explored, at least for what concerns the very early growth
stages.

When growing epitaxially on a substrate, the key factors involved in the stabilization of the film includes its cohesion energy, which is also affected by the strain induced by the lattice mismatch with the substrate, the interface energy, resulting from the chemical bonds formed between the film and the substrate, and the surface energy of the growing front exposed by the film (with or without strain). These are particularly relevant for a 2D island growth mode, in addition to a perimetral step energy that is really unknown in details and can be taken to be eventually negligible for large surface covering. In contrast, in the case of 3D island growth, the surface energy contribution of all exposed facets as well as the actual strain relaxation provided by the peculiar island shape should should play a major role.

In a recent paper \cite{Bertoni}, we presented a comparison of  volume and surface energies for $\mathrm{\alpha}$, $\mathrm{\beta}$, and $\mathrm{\kappa}$ phases of $\mathrm{Ga_2O_3}$, as calculated by density functional theory (DFT). We also accounted for the strain produced by the substrate and how it affects such results. For each phase, the elastic contribution of the lattice misfit with the sapphire substrate was calculated by imposing the matching between the planar network of oxygen atoms of the film and that of the substrate. Moreover, the elastic contribution of the lattice misfit with a plastically relaxed $\mathrm{\alpha-Ga_2O_3}$ buffer layer was considered. In fact, surveying the literature, one notices that $\mathrm{Ga_2O_3}$ phases can nucleate directly on the substrate or on a $\mathrm{\alpha-Ga_2O_3}$ interlayer, both relaxed or fully strained by a coherent interface with sapphire. \cite{schewski_ApplPhysExpr_2015}. The quantity missing in our previous work was the interface contribution that is the main topic of the present work. In Ref. \cite{porter} a thick film of $\mathrm{\kappa}$-$\mathrm{Ga_2O_3}$ is shown to grow on $\mathrm{\beta}$-$\mathrm{Ga_2O_3}$, possibly with some strain, then we also include the study of this interface in our work.

Here we demonstrate that the structural affinity between the film and the substrate (briefly, in terms of the percentage of octahedral and tetrahedral cages of O with respect to the fully octahedral structure of $\mathrm{\alpha-}$sapphire) is the leading criterion in lowering the interface energies and explains the quantitative meaning of the metastable phase-locking, especially the $\mathrm{\alpha}$ phase in mist-CVD.  

Finally, we address the following question: is the strain relaxation by dislocation nucleation in the $\mathrm{\alpha}$ phase disrupting the kinetic phase locking, or is it the nucleation and growth of 3D islands, as produced by higher temperatures and lower growth rates? By mist-CVD we know that the $\mathrm{\alpha}$ phase grows continuously in fully relaxed films \cite{Kaneko_2012}. Still, by MBE or MOCVD at different temperatures and rates, the $\mathrm{\beta}$ phase eventually appears. Therefore, we draw a simple nucleation model of the three phases considering 2D islands on sapphire, strained $\mathrm{\alpha}$ or relaxed $\mathrm{\alpha}$. The discussion below sheds some light on the experimental issues and provides some indications for future deposition tests.

\section*{Methods}

\subsection*{Interface energies calculations}

All the DFT calculations were performed using the VASP software \cite{vasp1, vasp2, vasp3}. We chose the Perdew–Burke–Ernzerhof exchange–correlation functional revised for solids (PBEsol)\cite{PBEsol}, maintaining continuity with our previous work on surface energies\cite{Bertoni}. We employed pseudo-potentials with 6 and 13 electrons in the valence states for O and Ga atoms, respectively. We optimized the atomic coordinates and the lattice parameters of all bulk structures, including the one of $\mathrm{\alpha}$-$\mathrm{Al_2O_3}$, using a plane-waves cutoff of 850 eV. All the resulting lattice parameters are reported in Table S1. Then, the following calculations, which do not require the optimization of lattice parameters, were performed with a reduced cutoff of 500 eV, in order to reduce the computational load. The Brillouin zone was sampled through (6 $\times$ 6 $\times$ 3), (2 $\times$ 12 $\times$ 6) and (7 $\times$ 4 $\times$ 4) unshifted Monkhorst–Pack k-point meshes for the $\mathrm{\alpha}$, $\mathrm{\beta}$ and $\mathrm{\kappa}$ cells, respectively.
We constructed the substrate slab by aligning its z axis with the c axis of sapphire. Therefore, the hexagonal in-plane lattice of the c-sapphire cell was our reference, with [100]/[120] directions along the x/y axes. 

The hetero-interfaces were modeled by stacking the slab of each film (i.e.  $\mathrm{\alpha-}$, $\mathrm{\beta-}$ and $\mathrm{\kappa-}$$\mathrm{Ga_2O_3}$) on the ones of different substrates: $\mathrm{\alpha-}$$\mathrm{Al_2O_3}$ and a fully strained/relaxed $\mathrm{\alpha-}$$\mathrm{Ga_2O_3}$ buffer layer. We also  considered a strained/relaxed interlayer of $\mathrm{\beta-}$$\mathrm{Ga_2O_3}$ in the case of $\mathrm{\kappa}$ film. This was done in a way that, again, film and the substrate share the layer of oxygen atoms at the interface. The slabs of the film were built in order to align the $\mathrm{\beta}$ [102] along the $\mathrm{\alpha}$ [100], the $\mathrm{\beta}$ [010] along the $\mathrm{\alpha}$ [120], the $\mathrm{\kappa}$ [100] along the $\mathrm{\alpha}$ [100] and the $\mathrm{\kappa}$ [010] along the $\mathrm{\alpha}$ [120]. When considering the interface between $\mathrm{\kappa}$ and $\mathrm{\beta}$  phases, the $\mathrm{\kappa}$ [100] is aligned along the $\mathrm{\beta}$ [102] and the $\mathrm{\kappa}$ [010] along the $\mathrm{\beta}$ [010]. Both slabs of the $\mathrm{Ga_2O_3}$ films and substrate were constructed in a way to expose one free surface, with surface structures corresponding to those investigated in our previous work\cite{Bertoni}. In order to prevent interactions between periodic replicas of each interface slab along its z-direction, we inserted a vacuum region of 13 \angstrom.
The interface slabs were built adapting the in-plane (super)cell paramaters of the film to that of the substrate. Thus, considering the epitaxial relationships between the film and the substrate, supercells replicating the slab unit cells of the film or substrate were exploited. The sizes of the supercells and the corresponding misfit strain applied to the film, for each interface, are detailed in Table S2. Finally, to obtain the relaxed interface structure, we optimized the atomic coordinates only, keeping the supercell fixed by the lattice constraints of the substrate. This approach allows a proper release of the stress along the z direction throughout the structural optimization.

As evident in Figure \ref{fgr:bulk}, all oriented $\mathrm{Ga_2O_3}$ phases can be considered as a stacking of bi-layers (a cation layer and an oxygen layer). When building the corresponding individual slabs, we converged the calculated interface energies with respect to the film and substrate thickness. Hence, when building the interface, we stacked slabs with 7 $\mathrm{\alpha}$-$\mathrm{Ga_2O_3}$, 9 $\mathrm{\beta}$-$\mathrm{Ga_2O_3}$ or 13 $\mathrm{\kappa}$-$\mathrm{Ga_2O_3}$ layers of Ga and O atoms (that we call bi-layers) on 7 $\mathrm{\alpha}$-$\mathrm{Al_2O_3}$ bi-layers.
Nonetheless, not all bi-layers are equivalen, since not all the planes show the same arrangement of Ga atoms with the same coordination (taking 2.4 \angstrom as cutoff for the first-neighbors distance). We then define the epitaxial layer as the unit that can be stacked to correctly generate a bulk. We mark the epitaxial layers in each phase in Figure \ref{fgr:bulk}. 

We calculated the interface energy as:
\begin{equation}\label{eq::gamma}
  \gamma_{int} = \frac{E_{slab} - \gamma_{sup}^{f}A - \gamma_{sup}^{s}A - N_{f}\mu_{f} - N_{s}\mu_{s} }{A} ,
\end{equation}
where $E_{slab}$ is the total energy of the interface slab, $\gamma^{f/s}_{sup}$ is the surface energy of the free bottom surface exposed by the substrate ($s$) or the top one film ($f$), with area $A$. $N_{s}$ and $N_{f}$ are the number of formula units contained in the substrate and film region of the slab, respectively. $\mu$ is the bulk chemical potential: it accounts also for the elastic energy contribution, if any strain is applied. Surface and bulk energies were re-calculated with the current setup, still using the same methodology as in our previous work\cite{Bertoni}.

\section*{Results and Discussion}

\subsection*{Interface energies}

\begin{table*}[h!]
\small
  \caption{Interface energies (in meV/\angstrom$^2$) for different substrate/film combinations. The label $\varepsilon$ marks those $\mathrm{Ga_2O_3}$ interlayers strained to fit the lattice parameters of c-sapphire. }
  \label{tbl:gammaresults}
  \begin{tabular*}{\textwidth}{@{\extracolsep{\fill}}llllll}
    \hline
    film & $\mathrm{\alpha-Al_2O_3}$ & $\mathrm{\alpha-Ga_2O_3}$ & $\mathrm{\alpha-Ga_2O_3^{\varepsilon}}$ & $\mathrm{\beta-Ga_2O_3}$ & $\mathrm{\beta-Ga_2O_3^{\varepsilon}}$ \\
    \hline
    $\mathrm{\alpha-Ga_2O_3}$ & -3 & - & - & - & - \\
    $\mathrm{\beta-Ga_2O_3}$  & 53 & 47 & 52 & - & - \\
    $\mathrm{\kappa-Ga_2O_3}$ & 27 & 32 & 30 & 21 & 15 \\
    \hline
  \end{tabular*}
\end{table*}

To provide additional insight into the understanding the epitaxial growth mechanism of different $\mathrm{Ga_2O_3}$ phases on c-sapphire, we calculated such interface energies $\gamma_{int}$. Since some experimental works indicate the presence of an intermediate $\mathrm{Ga_2O_3}$ layer of a different phase, which can be an $\mathrm{\alpha}$ interlayer, coherent \cite{schewski_ApplPhysExpr_2015} or relaxed \cite{growth_dir5}, for the $\mathrm{\beta}$ phase and for the $\mathrm{\kappa}$ phase \cite{growth_dir3}, or a $\mathrm{\beta}$ interlayer for the $\mathrm{\kappa}$ phase \cite{porter}, we also calculated $\gamma_{int}$ in these cases. For the sake of simplicity, our calculations were focused on sharp interfaces only, i.e. with no intermixing of cation species across the interface. Indeed, such an assumption may constitute a simplification of the actual interfaces \cite{Schowalter}, but allows us to provide a more comprehensible interpretation of the physical effects that affect the stabilization of the interface. Finally, despite the fact that rotational domains of 120° are known to occur for both $\mathrm{\beta}$ and $\mathrm{\kappa}$ films on sapphire, we calculated interface energies for just one (equivalent) domain orientation, as it is still not understood whether some strain release can be introduced at domain boundaries.

The structure of the stacked bi-layers is identical in the case of $\mathrm{\alpha}$-$\mathrm{Ga_2O_3}$ on sapphire (but for an in-plane shift); therefore, only one possible termination can be found matching the position of O atoms at the interface. On the contrary, both in $\mathrm{\beta}$ and $\mathrm{\kappa}$ bulk cells two different bi-layers are alternated (see Figure \ref{fgr:bulk}b,c). Each of them has a different alternation of octahedral (or pentahedral) and tetrahedral Ga atoms, leading to either two or four different types of interfaces. In Table \ref{tbl:gammaresults} we collect only the interface configurations that lead to the lowest energies. In the case of $\mathrm{\beta}$ and $\mathrm{\kappa}$ films on sapphire, the additional results calculated with other terminations are collected in Figure S1-2.

Generally speaking, these values of $\gamma_{int}$ are sensibly lower than the surface energies\cite{Bertoni}, as expected. Still, as in the case of the surface energies, it is quite useful to interpret our data by performing an analysis of the optimized configurations through the number of first neighbors of Ga atoms, assuming a cut-off distance of 2.4 \angstrom.

\begin{figure}[h]
\centering
  \includegraphics[height=4.9cm]{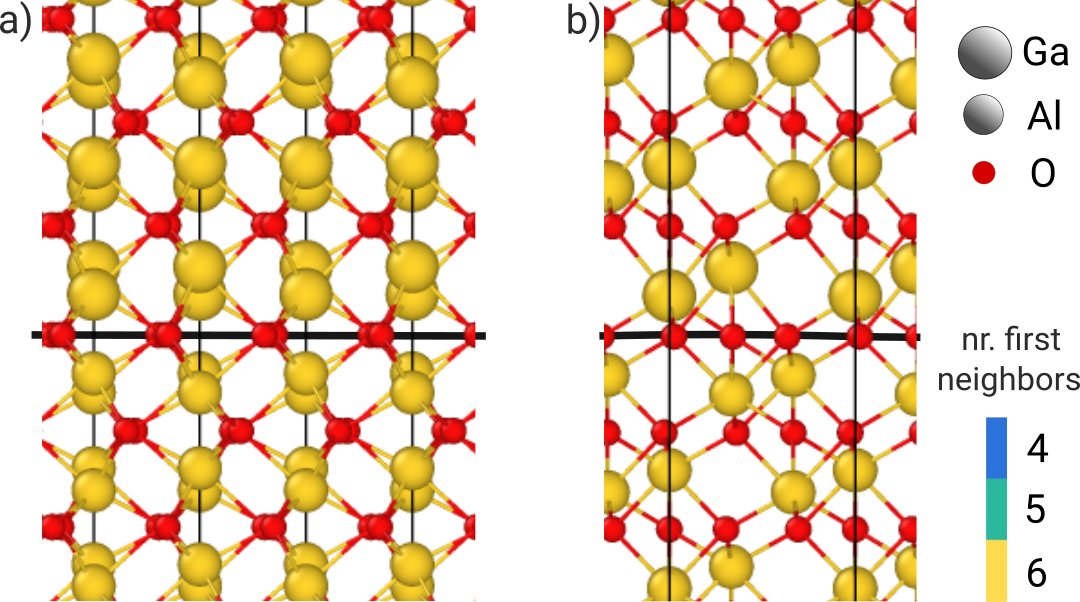}
  \caption{Front (a) and side (b) view of the $\mathrm{\alpha-Ga_2O_3}$/$\mathrm{\alpha-Al_2O_3}$ interface. The black line marks  the plane of O atoms shared by both film and substrate.}
  \label{fgr:Int a/a-Al2O3}
\end{figure}

In the case of $\mathrm{\alpha}$-$\mathrm{Ga_2O_3}$, the interface with $\mathrm{\alpha}$-$\mathrm{Al_2O_3}$ , reported in Figure \ref{fgr:Int a/a-Al2O3}, is almost homogeneous, as the two slabs have the same structure. No distortion is found nor deviation from the ideal six-fold coordination of the cations: therefore, the corresponding interface behaves almost as a continuation of the bulk. In fact, the calculated interface energy turns out to be -3 meV/ \angstrom$^2$ and can be considered negligible, within the precision of our calculation method.

\begin{figure}[h]
\centering
  \includegraphics[height=5.2cm]{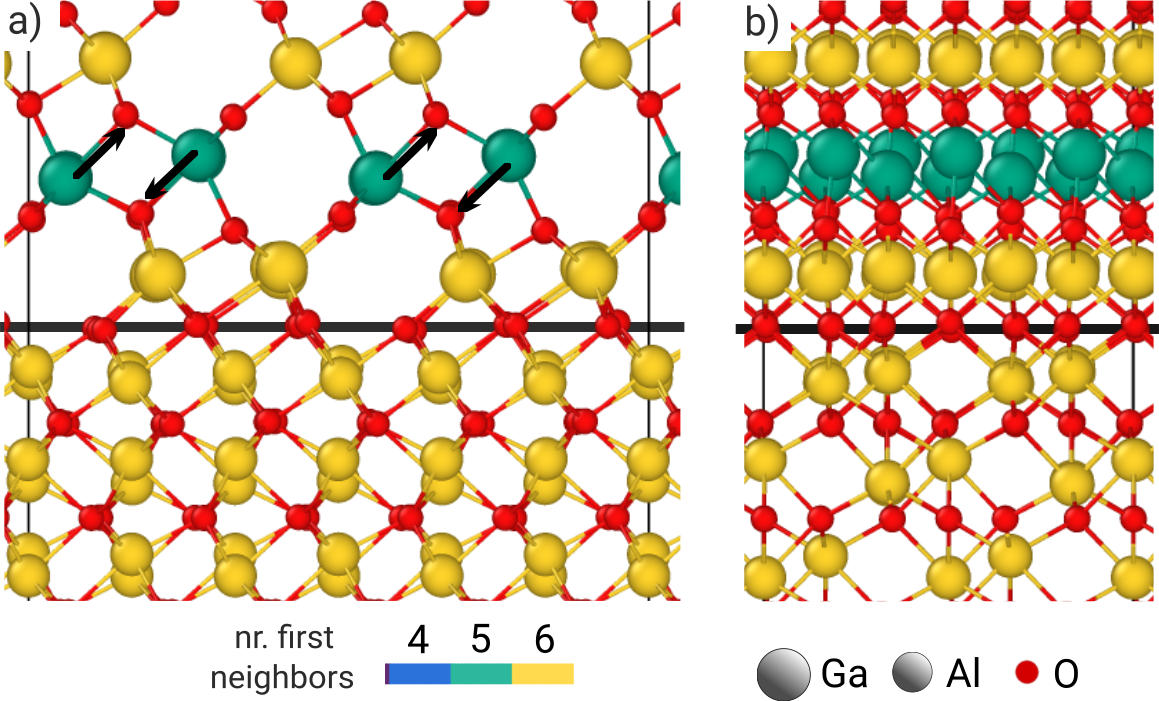}
 \caption{Front (a) and side (b) view of the $\mathrm{\beta-Ga_2O_3}$/$\mathrm{\alpha-Al_2O_3}$ interface. The black line marks  the plane of O atoms shared by both film and substrate.}
  \label{fgr:Int b/a-Al2O3}
\end{figure}

In contrast, some change in coordination is expected for  $\mathrm{\beta}$-$\mathrm{Ga_2O_3}$, when forming the interface with sapphire, as the interface energy is much higher. In fact, each tetrahetrally coordinated Ga atom belonging to the second row above the interface gains one additional O neighbor each (see the black arrows in Figure \ref{fgr:Int b/a-Al2O3}). This change is related to how the Ga and O atoms settle in the first two rows above the interface. The structural and energetic optimization appears not to be affected by the strain, nor by the atomic species of the cation in the substrate, as the same change in coordination is found for the strained and the fully relaxed $\mathrm{\alpha}$-$\mathrm{Ga_2O_3}$ interlayer as well (see in Figure S3-S4). Hence, the difference in structural environment for cations at the interface leads to a sizeable interface energy (as reported in Table 1), that is almost the same for $\mathrm{\alpha}$-$\mathrm{Ga_2O_3}$ fully strained/relaxed substrate. This trend is confirmed when considering the second interface between $\mathrm{\beta}$ phase and sapphire (see Figure S1), which displays an array of tetrahedrally coordinated cations at the interface in place of octahedrally coordinated ones: the surface energy is higher (77 meV/\angstrom $^2$) despite the optimization turns most interface cations to a pentahedral coordination.

\begin{figure}[h]
\centering
  \includegraphics[height=4.8cm]{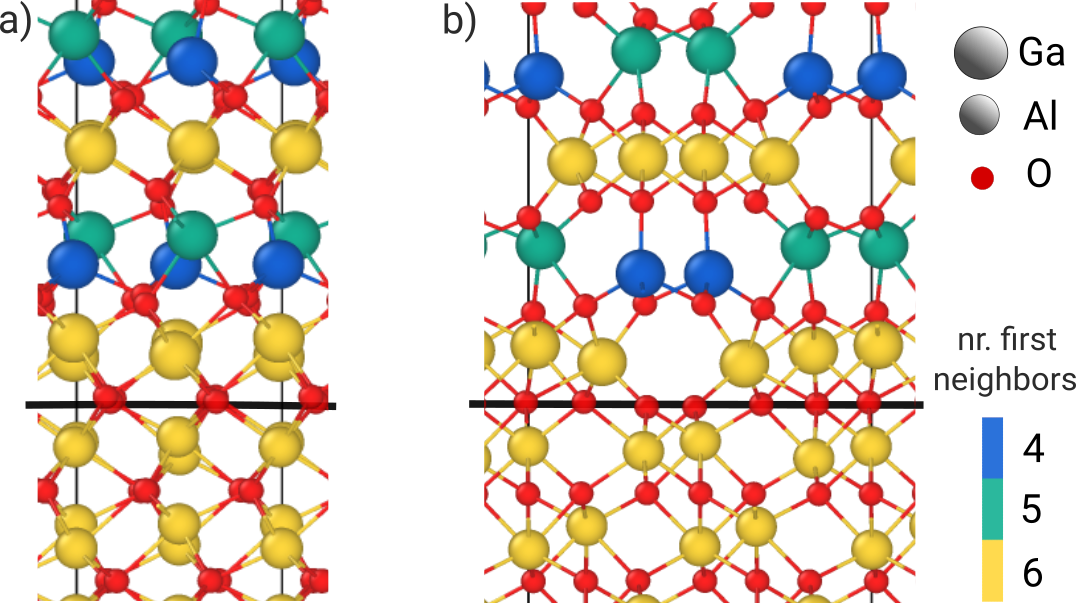}
  \caption{Front (a) and side (b) view of the $\mathrm{\kappa-Ga_2O_3}$/$\mathrm{\alpha-Al_2O_3}$ interface. The black line marks  the plane of O atoms shared by both film and substrate.}
  \label{fgr:Int k/a-Al2O3}
\end{figure}

When considering $\mathrm{\kappa}$-$\mathrm{Ga_2O_3}$ interface with sapphire, reported in Figure \ref{fgr:Int k/a-Al2O3}, no change in coordination for cations with respect to their bulk occurs, neither for the substrate nor for the film. It is fair to say that in this case the $\mathrm{\kappa}$ phase displays at the interface a layer of cations already in the octahedral configuration, and the interface energy is actually low, about half the one of $\mathrm{\beta}$ on sapphire. The same structure is found when an $\mathrm{\alpha}$-$\mathrm{Ga_2O_3}$ buffer layer is present, either strained or fully relaxed (see Figure S5-S6), and its interface energy is roughly the same as with sapphire. In case the second possible interface with sapphire is considered (see Figure S2), the interface energy raises to 38 meV/\angstrom$^2$, as in this case a layer of cations in tetrahedral and pentahedral coordination is present at the interface, and the structural optimization drives a tendency to turn them to pentahedral and octahedral coordination, respectively.

\begin{figure}[h]
\centering
  \includegraphics[height=5.9cm]{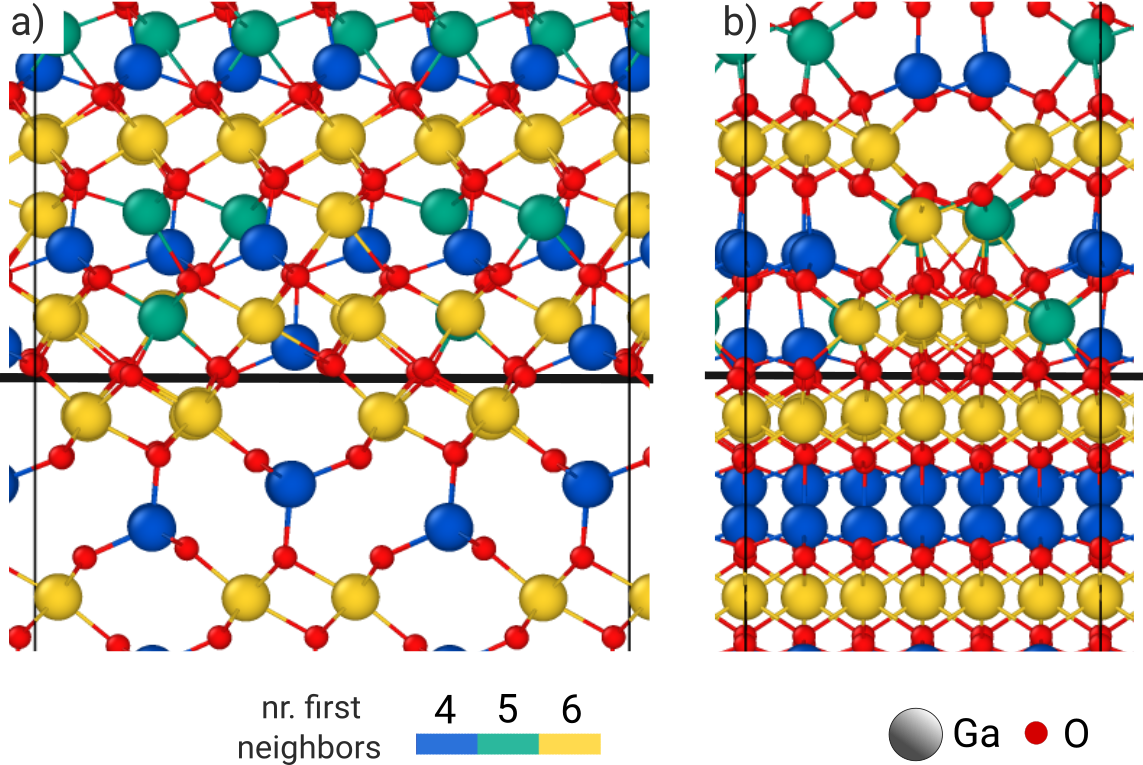}
  \caption{Front (a) and side (b) view of the $\mathrm{\kappa-Ga_2O_3}$/$\mathrm{\beta-Ga_2O_3}$ interface. $\mathrm{\beta-Ga_2O_3}$ interlayer is fully strained on $\mathrm{\alpha-Al_2O_3}$ substrate. The black line marks the plane of O atoms shared by both film and substrate.}
  \label{fgr:Int k/b-Ga2O3}
\end{figure}

Interestingly, the case of an interface between the two most open structures, i.e. $\mathrm{\kappa}$ phase over either a fully strained or fully relaxed $\mathrm{\beta}$ interlayer (represented in Figure \ref{fgr:Int k/b-Ga2O3} and Figure S7), results in a very low interface energy. This is the case especially for a strained $\mathrm{\beta}$ phase interlayer. Here, out of the four possible configurations, we are considering the one in which both the film and the substrate display the same coordination for cations at the interface, i.e. in octahedral cages: this is a starting point with a presumable low interface energy, but it further decreases with the structural optimization. This can be explained by the fact that the more open interface allows to turn some more distant cations of the $\mathrm{\kappa}$ phase with octahedral coordination to the fourfold ones, which seems to mimic the $\mathrm{\beta}$ stacking. 

It is therefore quite evident that out of our result a few clear trends emerge. 1) the structures forming an interface display a lower interface energy if the structures are more similar in the percentage of cations in tetrahedral and octahedral coordination. $\mathrm{\alpha-Ga_2O_3}$ on sapphire being the lowest in energy, $\mathrm{\beta}$ on sapphire the highest in energy. 2) Particularly, the similarity in structure affects the interface region, so that those configurations preserving a smoother change in coordination are lower in energy, as in the case of interfaces shown in Figure 2-4 and S3-S6, with respect to their alternative interface terminations reported in Figure S1-S2. 3) The structural optimization is effective in doctoring the change in coordination across the interface, lowering the interface energy, specifically for more open structures, such as $\mathrm{\beta}$ and $\mathrm{\kappa}$ phases. 

\subsection*{Energetics of $\mathrm{Ga_2O_3}$ epitaxial layers and phase locking}
The calculations of interfacial energies from the previous section, combined with the data of surface energy and strain computed in Ref.~\cite{Bertoni} and reported in Table S3, allows for a quantitative evaluation of the relative stability of $\mathrm{Ga_2O_3}$ epitaxial layers for each phase on either $\mathrm{\alpha-Al_2O_3}$ or $\mathrm{\alpha-Ga_2O_3}$ substrates. From classical nucleation theory, the formation energy of a 2D island of phase $p$ can be expressed as a function of the number $n$ of formula units:
\begin{equation}\label{eq::deltaG_gen}
    \Delta G^p(n) = -\left(\mu_\mathrm{gas}-\mu_\mathrm{bulk}^p - \Delta\mu_\mathrm{\epsilon}^p\right)n + \left(\gamma_\mathrm{epi}^p + \gamma_\mathrm{int}^p - \gamma_\mathrm{sub}\right) A^p(n) + \Lambda^p L^p(n)
\end{equation}
The terms in the leftmost brackets account for the energy balance between the gas and the solid phase: $\mu_\mathrm{gas}-\mu_\mathrm{bulk}^p$ is the gas supersaturation relative to the (relaxed) bulk $\mathrm{Ga_2O_3}$ in $p$ phase, while $\Delta\mu_\mathrm{\epsilon}^p$ is the elastic energy density within the 2D island, due to the lattice mismatch between the film and the substrate. The second group of terms accounts for the net change in surface energy density when covering an area $A^p$ of the substrate, displaying a surface energy $\gamma_\mathrm{sub}$, with the island.  $\gamma_\mathrm{epi}^p$ and $\gamma_\mathrm{int}^p$ are the corresponding energy densities of of the film free-surface and its interface energy, respectively. We explicit the dependency on the number of formula units as we define $A^p(n)=v^p/h^p*n$ with $v^p$ the volume per $p-\mathrm{Ga_2O_3}$ unit-formula and $h^p$ the height of the epitaxial layer as defined in the Methods section. In Table S3 we report also the $v_{p}$ and $h_{p}$ values, along with the surface energy of the sapphire (0001) substrate. Finally, the third term accounts for the energy cost of island edges, proportional to the length of its perimeter $L^p\propto\sqrt n$, by a coefficient $\Lambda^p$, determined by the actual island shape and by the, orientation-dependent, linear energy density of its edges, which are both out of our knowledge. 

Since we are interested only in the comparison of the stability  between the phases, we conveniently rephrase the Eqn.~\eqref{eq::deltaG_gen} to obtain the net energy density (per unit-formula) of a 2D $\mathrm{Ga_2O_3}$ layer in $p$ phase as
\begin{equation}\label{eq::deltaG_gen_b}
    \rho_\mathrm{2D}^p = \frac{\Delta G^p - \mu_\mathrm{ref}}{n} =\mu_\mathrm{bulk}^p + \Delta\mu_\mathrm{\epsilon}^p +\frac{v^p}{h^p} \Delta\gamma^p
\end{equation}
where $\Delta\gamma^p=\gamma_\mathrm{epi}^p + \gamma_\mathrm{int}^p - \gamma_\mathrm{sub}$ and $\mu_\mathrm{ref}=-\mu_\mathrm{gas}n + \Lambda\sqrt{n}$. In the latter we include the edge contribution and the reason is two-fold. We assume on statistical grounds that the differences between $\mathrm{Ga_2O_3}$ islands of different phases are negligible; moreover, we will focus on mature growth stages with large $n$, hence we deem the perimeter contribution less important.

\begin{figure}[h]
\centering
  \includegraphics[height=6.cm]{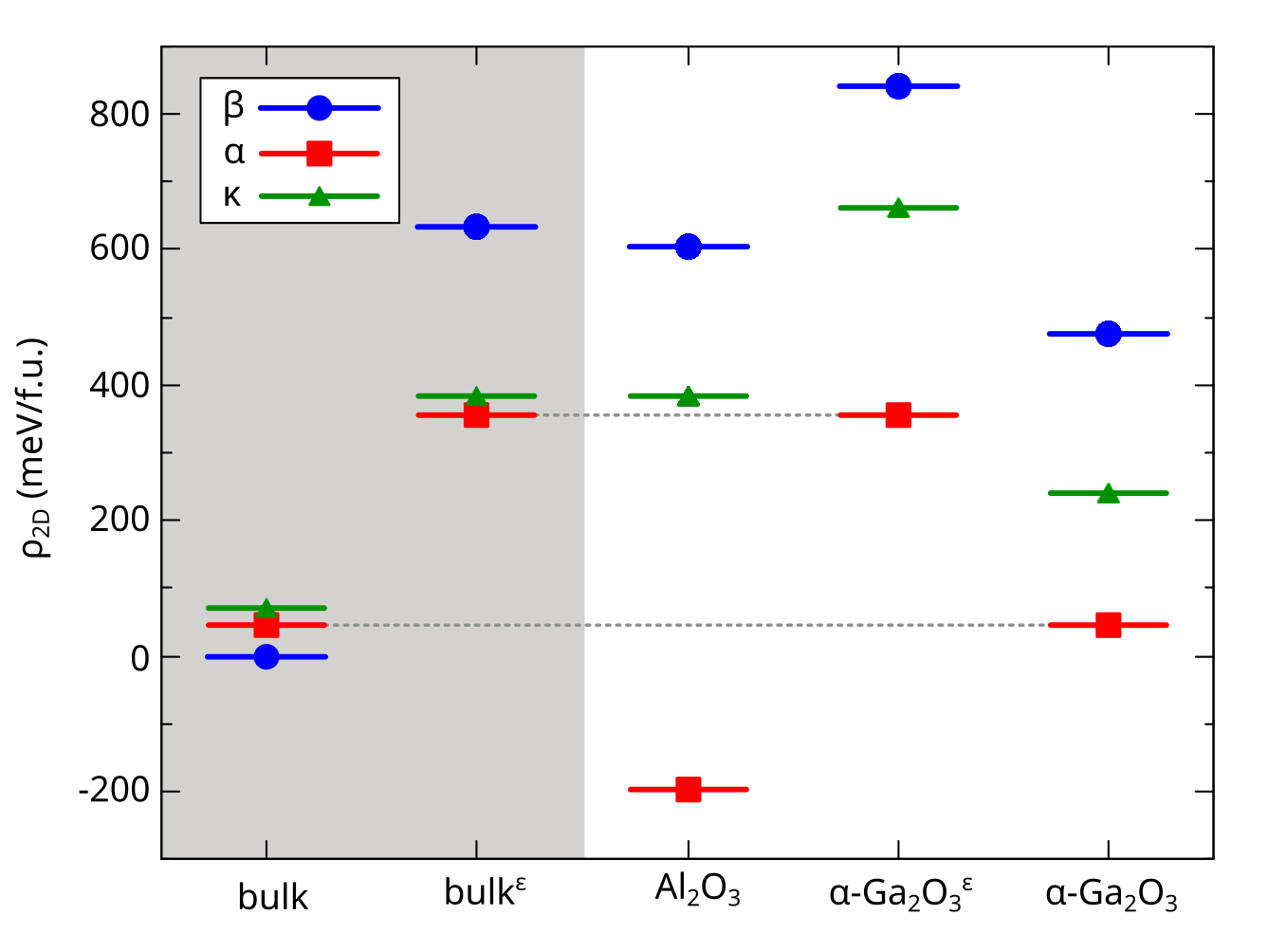}
  \caption{Diagram of $\rho_{2D}$ calculated for 2D films of the $\mathrm{Ga_2O_3}$ polymorphs, considering different substrates. The bulk values are reported on the left as reference, also accounting for the misfit strain induced by the $\mathrm{Al_2O_3}$ substrate (labelled with $^{\varepsilon}$). }
  \label{fgr:energylevels}
\end{figure}

The volumetric and surface contributions calculated for 2D layers of each $\mathrm{Ga_2O_3}$ phase on either $\alpha-Al_2O_3$ and $\alpha-Ga_2O_3$, both strained to sapphire lattice parameter and fully relaxed, are reported in Table~\ref{tbl:energylayer}. The resulting $\rho_\mathrm{2D}$ values are reported as energy levels in Figure ~\ref{fgr:energylevels}.

\begin{table*}[t]
\small
  \caption{Volumetric and surface/interface contributions of Eqn.~\ref{eq::deltaG_gen_b} reported in $\mathrm{meV/f.u.}$ for 2D layers of the three phases of $\mathrm{Ga_2O_3}$ on different substrate. The bulk energy of $\beta-\mathrm{Ga_2O_3}$ is taken as reference.  The label $\varepsilon$ marks those $\mathrm{Ga_2O_3}$ substrates strained to fit the lattice parameters of c-sapphire. }
  \label{tbl:energylayer}
  \begin{tabular*}{\textwidth}{@{\extracolsep{\fill}}llcccccc}
    & \multicolumn{1}{r}{substrate:} & \multicolumn{2}{c}{$\mathrm{\alpha-Al_2O_3}$} & \multicolumn{2}{c}{$\mathrm{\alpha-Ga_2O_3^{\varepsilon}}$} & \multicolumn{2}{c}{$\mathrm{\alpha-Ga_2O_3}$} \\
     \cline{3-8}
     & $\mu_\mathrm{bulk}$ & $\Delta\mu_\mathrm{\varepsilon}$ & $(v/h)\Delta\gamma$ & $\Delta\mu_\mathrm{\varepsilon}$ & $(v/h)\Delta\gamma$ &$\Delta\mu_\mathrm{\varepsilon}$ & $(v/h)\Delta\gamma$ \\
    \hline
    $\mathrm{\alpha-Ga_2O_3}$ & 47  & 309 & -553 & 309 & 0  & 0   & 0  \\
    $\mathrm{\beta-Ga_2O_3}$  & 0 & 633 & -30  & 633 & 207 & 150 & 325 \\
    $\mathrm{\kappa-Ga_2O_3}$ & 65 & 314 & 0    & 314 & 276 & 10  & 163 \\
    \hline
  \end{tabular*}
\end{table*}

While for the bulk phases (relaxed, first column in the gray area) the most stable phase is $\beta-\mathrm{Ga_2O_3}$, followed by $\alpha$ and $\kappa$, when growing epitaxial layers on $\alpha-\mathrm{Al_2O_3}$ such ordering is radically changed. Indeed, the lattice mismatch of a single domain with the substrate results in a biaxial strain ($\epsilon$), generating a substantial increment of the bulk chemical potential as made evident in the second column of Figure~\ref{fgr:energylevels}, for the infinite bulk correspondingly strained. The effect is more dramatic for the $\beta$ phase, which becomes strongly unfavorable even against the $\kappa$ one. In the third column, the surface energy of sapphire substrate comes in, along with its substitution by the interface energy and the surface energy of the first $\mathrm{Ga_2O_3}$ epitaxial layer on the substrate. Here, the energy levels of $\beta$ and $\kappa$ phases are nearly aligned with the second column, as $\Delta\gamma\approx0$. In contrast, for the first epitaxial layer of $\alpha-\mathrm{Ga_2O_3}$ the surface energy gain from replacing the costly (001) $\alpha-\mathrm{Al_2O_3}$ (113 meV/\angstrom$^2$) surface with the more convenient (001) $\alpha-\mathrm{Ga_2O_3}$ (88 meV/\angstrom$^2$) , along with the negligible benefit in forming the interface, results in a negative $\Delta\gamma$ that largely compensate the elastic energy. This surface/interface advantage is so large that growing the $\alpha-\mathrm{Ga_2O_3}$ epilayer on $\alpha-\mathrm{Al_2O_3}$ becomes even more convenient than the formation of the bulk phase. 

In the fourth and fifth columns, we find the values for one epitaxial layer on top a coherently strained interlayer of $\alpha-\mathrm{Ga_2O_3}$, respectively. As adding one epitaxial layer of the $\alpha$ phase on top is like to add one layer inside the (thick) intralayer, the energy values of the former case correctly aligns to the bulk value with the corresponding strain, while the latter one correctly aligns to the relaxed bulk value (horizontal dashed lines). The hierarchy with respect to the $\beta$ and $\kappa$ phase does not change substantially and it is important to say that such situation eventually corresponds to the additional layers after the very first few ones, either coherent to the substrate, or with a plastic relaxation of the alpha phase.

These results explain the phase locking mechanism proposed by Kaneko\cite{growth4} for the $\alpha$ phase in mist-CVD, because the hierarchy of the stability of different phases of the film follows the structural similarity with the substrate. In this respect, the strongest driving force is the lattice parameter affinity, i.e. the elastic energy contribution, but the affinity in the octahedral and tetrahedral configuration plays the main role at reduced strain. Therefore, our model supports the epitaxial growth of $\alpha-\mathrm{Ga_2O_3}$ film first on the bare c-sapphire substrate, either as a thin interfacial layer\cite{schewski_ApplPhysExpr_2015} or as a thick crystal\cite{Kaneko_2012}. Therefore, it appears that the only way to explain the growth of the $\beta$  phase straightforwardly on a sapphire substrate, or after an  $\alpha$ phase interlayer, is to allow for elastic relaxation by 3D island nucleation, or to some plastic relaxation induced by domain borders, more likely grain boundaries among coalesced 3D islands, or columnar grains. In both cases, higher growth temperatures and reduced growth rates are necessary, in order to allow for larger surface mean free paths. Unfortunately, the morphology of the very early stages of  $\beta$  and  $\kappa$  phase growth and the corresponding strain release have not been characterized so far to a sufficient extent to draw an accurate modelling, including such kinetic effects. 
\section*{Conclusions}
In this work we calculated the interface energies for $\alpha$, $\beta$ and $\kappa$ phases of $\mathrm{Ga_2O_3}$ on c-sapphire substrate by considering the presence of a shared network of O atoms between film and substrate. We investigated the role of the strain at the interface and we identified the affinity in the percentage of cations coordination between the two sides of the interface as the major driving force for the lowest energies.

Using these values, we then performed a comparison between the relative stability of epitaxial films of the $\mathrm{Ga_2O_3}$ polymorphs. Our analysis reveals that for purely 2D growth of $\mathrm{Ga_2O_3}$ on $\alpha-\mathrm{Al_2O_3}$ the $\alpha$ phase is the most favorable and remains as such even if considering the nucleation of a new layer on top of an existing, either relaxed or not, $\mathrm{\alpha-Ga_2O_3}$ interlayer. This provides a quantitative interpretation of the phase locking effect proposed in Ref.~\cite{growth4}, corroborating the idea that $\alpha-\mathrm{Ga_2O_3}$ is preferred in mist-CVD experiments, due to its convenient interaction with the underlying substrate.

\section*{Author contributions}
I.B. investigation, visualization, writing-original draft; A.U. resources, data curation, writing-reviewing\&editing; E.S. validation, methodology, writing-reviewing\&editing; R.B. investigation, methodology, writing-reviewing\&editing; L.M. conceptualization, funding acquisition, supervision.

\section*{Conflicts of interest}
There are no conflicts to declare.

\section*{Data availability}

All details to reproduce the calculations are reported in the Methods Section of the main manuscript and the files of all the optimized structures are collected in the archive supplementary\_information\_geometries.zip .

\section*{Acknowledgements}

This study was carried out within the MOST–Sustainable Mobility Center and received funding from the European Union Next-GenerationEU (PIANO NAZIONALE DI RIPRESA E RESILIENZA (PNRR)–MISSIONE 4 COMPONENTE 2, INVESTIMENTO 1.4 – D.D. 1033 17/06/2022, CN00000023).
We acknowledge the CINECA consortium under the ISCRA initiative for the availability of high-performance computing resources and support.

\balance

\renewcommand\refname{References}

\bibliography{rsc} 
\bibliographystyle{rsc} 
\end{document}


\centering
\LARGE{SUPPLEMENTARY INFORMATION FOR}\\
\vspace{1cm}
\Large{Interface energies of $\mathrm{Ga_2O_3}$ phases with the sapphire substrate and the phase-locked epitaxy of metastable structures explained}\\
\vspace{1cm}
\normalsize{Ilaria Bertoni${\ast}^{a}$, Aldo Ugolotti$^{a}$, Emilio Scalise$^{a}$, Roberto Bergamaschini$^{a}$, and Leo Miglio$^{a}$}\\
\vspace{0.5cm}
\small{\textit{$^{a}$~Department of Materials Science, University of Milano-Bicocca, via Cozzi 55, 20125 Milan (Italy)}}\\
\vspace{3.5cm}
\begin{table}[h]
    \centering
    \begin{tabular}{|c|c|c|c|} \hline 
         phase&  a[\angstrom]&  b[\angstrom]&  c[\angstrom] \\ \hline
         $\mathrm{\kappa-Ga_2O_3}$ & 5.061 & 8.686 &  9.292  \\ \hline
         $\mathrm{\beta-Ga_2O_3}$ & 14.763 & 3.048  &  5.809  \\ \hline 
         $\mathrm{\alpha-Ga_2O_3}$ & 5.001 & 5.001 & 13.448  \\ \hline
         $\mathrm{\alpha-Al_2O_3}$ & 4.775 & 4.775 & 13.015  \\ \hline 
    \end{tabular}
    \caption{Lattice parameters of optimized bulk cells, oriented according to the substrate. The slabs of the film were built in order to align the $\mathrm{\beta}$ [102] along the $\mathrm{\alpha}$ [100], the $\mathrm{\beta}$ [010] along the $\mathrm{\alpha}$ [120], the $\mathrm{\kappa}$ [100] along the $\mathrm{\alpha}$ [100] and the $\mathrm{\kappa}$ [010] along the $\mathrm{\alpha}$ [120].}
    \label{tab:my_label}
\end{table}

\begin{table}
    \centering
    \begin{tabular}{|c|c|c|c|c|} \hline 
         \multicolumn{2}{|c}{} & \multicolumn{3}{|c|}{m$_x$ / m$_y$ misfit strain}  \\ \hline
         phase&  x/y axis& vs $\mathrm{\alpha-Al_2O_3}$& vs $\mathrm{\alpha-Ga_2O_3}$ & vs $\mathrm{\beta-Ga_2O_3}$\\ \hline
         $\mathrm{\kappa-Ga_2O_3}$ &  [100]/[010]& 4.8 \% / 5.7 \% & 0.3 \% / -1.2 \% & -5.3 \% / 2.8 \% \\ \hline
         $\mathrm{\beta-Ga_2O_3}$ &  [102]/[010]& 9.5 \% / 3.0 \% & 5.3 \% / -1.6 \% &-\\ \hline 
         $\mathrm{\alpha-Ga_2O_3}$ &  [100]/[120]& 4.5 \% / 4.5 \% & - &-\\ \hline
         $\mathrm{\alpha-Al_2O_3}$ &  [100]/[120]& - & -  &-\\ \hline 
        \multicolumn{2}{|c}{} & \multicolumn{3}{|c|}{epitaxial relationship: n$_x$$\times$n$_y$ (film): n$_x$$\times$n$_y$ (substrate)}  \\ \hline
         phase&  in-plane cell& vs $\mathrm{\alpha-Al_2O_3}$& vs $\mathrm{\alpha-Ga_2O_3}$ & vs $\mathrm{\beta-Ga_2O_3}$\\ \hline
         $\mathrm{\kappa-Ga_2O_3}$ &  conventional-rectangular & 1$\times$1:1$\times$1 & 1$\times$1:1$\times$1 & 1$\times$1:1$\times$1 \\ \hline
         $\mathrm{\beta-Ga_2O_3}$ &  conventional-rectangular& 1$\times$3:3$\times$1 & 1$\times$3:3$\times$1 & 1$\times$3:3$\times$1\\ \hline 
         $\mathrm{\alpha-Ga_2O_3}$ &  primitive-hexagonal& 1$\times$1:1$\times$1 & 1$\times$1:1$\times$1 &1$\times$1:1$\times$1\\ \hline
    \end{tabular}
    \caption{Misfit strain with the substrate, calculated as: m$_i$=$(a_{i}^{film} - a_{i}^{substr})/a_{i}^{film}$. $a_i$ is the lattice parameter of the film/substrate along the given direction.}
    \label{tab:my_label}
\end{table}


\begin{table}
    \centering
    \begin{tabular}{|c|c|c|c|c|}\hline 
           phase & $\mathrm{$\Delta\mu_\mathrm{\varepsilon}$} $ [meV/f.u.]& $\mathrm{\gamma_{epi}}$  [meV/f.u.] & $v $ [\angstrom^{3}/f.u.]& $h$  [\angstrom]\\ \hline 
         \multicolumn{5}{|c|}{on $\mathrm{\alpha-Al_2O_3}$ substrate}\\ \hline 
         $\mathrm{\alpha}$ & 309 & 88	& 45.8 & 2.32 \\ \hline 
         $\mathrm{\beta}$& 633& 57& 58.7&5.94\\ \hline 
         $\mathrm{\kappa}$ &  314& 86 & 48.1 & 4.88 \\ \hline
         \multicolumn{5}{|c|}{on $\mathrm{\alpha-Ga_2O_3^{\varepsilon}}$ substrate}\\\hline
         $\mathrm{\alpha}$ & 309 & 88	& 45.8 & 2.32 \\ \hline 
         $\mathrm{\beta}$& 633& 57& 58.7&5.94\\ \hline 
         $\mathrm{\kappa}$ & 314  & 86 &  48.1 & 4.88 \\ \hline
         \multicolumn{5}{|c|}{on $\mathrm{\alpha-Ga_2O_3}$ substrate}\\\hline
         $\mathrm{\alpha}$ & 0 & 70 & 48.5 & 2.24 \\ \hline 
         $\mathrm{\beta}$& 150 & 53 & 51.3 & 4.74 \\ \hline 
         $\mathrm{\kappa}$ &  10 & 53 & 50.8 & 4.69 \\ \hline
         \multicolumn{5}{|c|}{  } \\\hline
         \multicolumn{5}{|c|}{$\mathrm{\alpha-Al_2O_3}$:    ${\gamma_{epi}=113meV/\angstrom^2}$} \\\hline
    \end{tabular}
    \caption{Elastic energy, surface energy of the epilayer, atomic volume of the cell and thickness of the layer for the different phases, given $\mathrm{\alpha-Al_2O_3}$ substrate, $\mathrm{\alpha-Ga_2O_3}$ strained interlayer and $\mathrm{Ga_2O_3}$ fully relaxed interlayer.}
    \label{tab:my_label}
\end{table}

\newpage

\begin{figure}[h]
\centering
  \includegraphics[height=5.2cm]{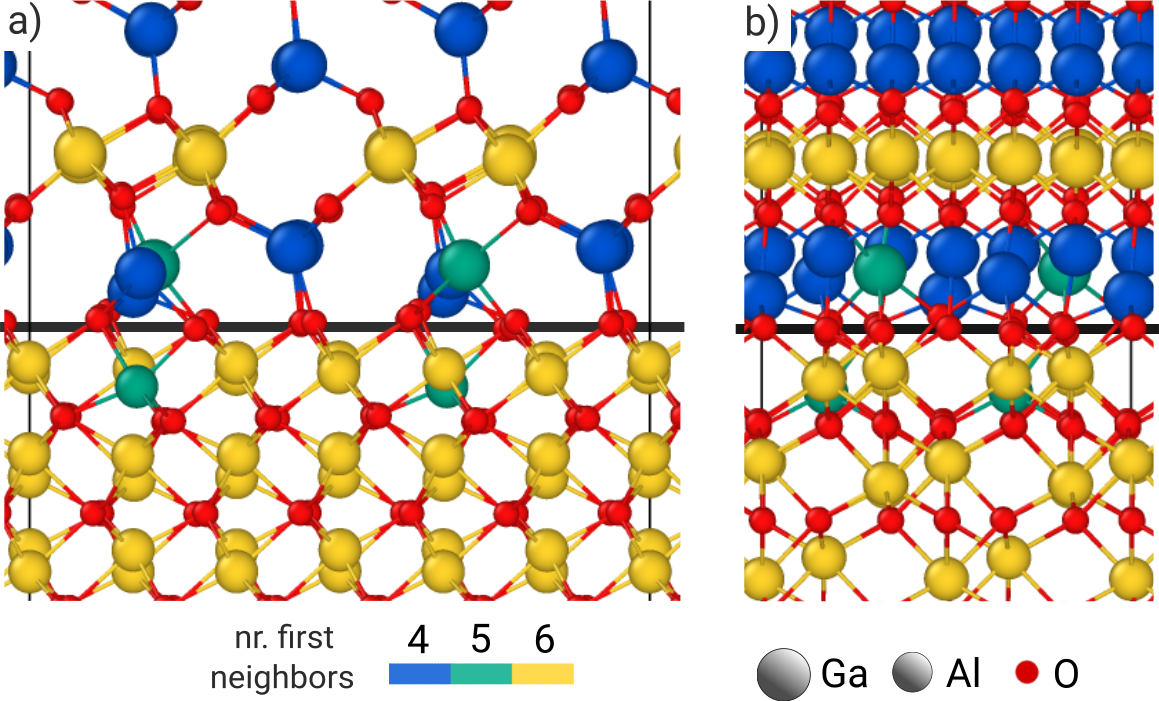}
 \caption{\label{fgr:Int b/a-Al2O32} Front (a) and side (b) view of the $\mathrm{\beta-Ga_2O_3}$/$\mathrm{\alpha-Al_2O_3}$ interface nr. 2. The black line marks  the plane of O atoms shared by both film and substrate.}
\end{figure}

\begin{figure}[h]
\centering
  \includegraphics[height=5.2cm]{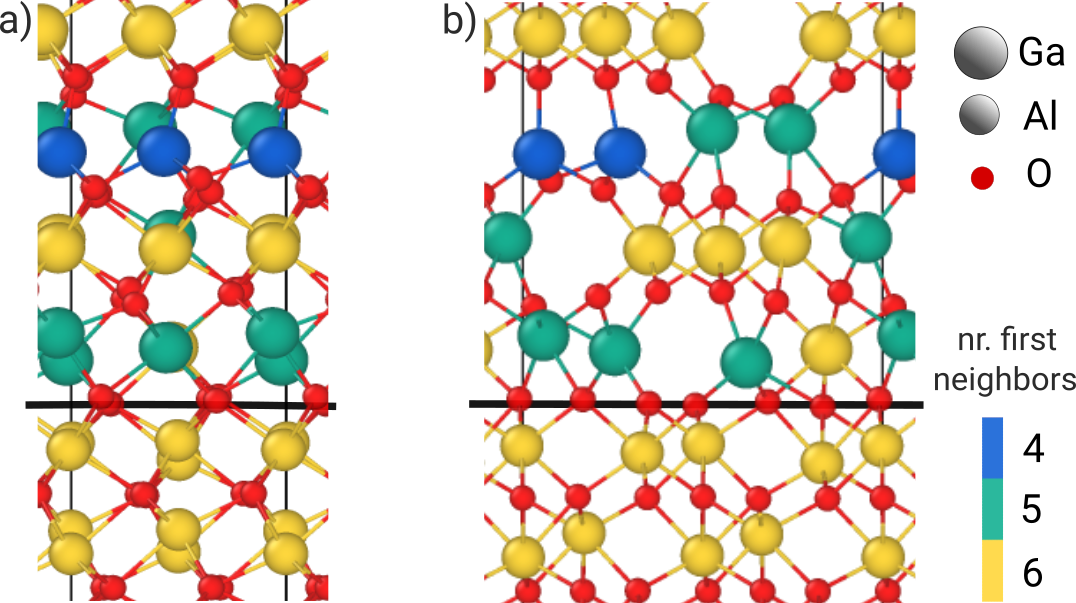}
 \caption{\label{fgr:Int k/a-Al2O3}Front (a) and side (b) view of the $\mathrm{\kappa-Ga_2O_3}$/$\mathrm{\alpha-Al_2O_3}$ interface nr. 2. The black line marks  the plane of O atoms shared by both film and substrate.}
\end{figure}

\begin{figure}[h]
\centering
  \includegraphics[height=5.2cm]{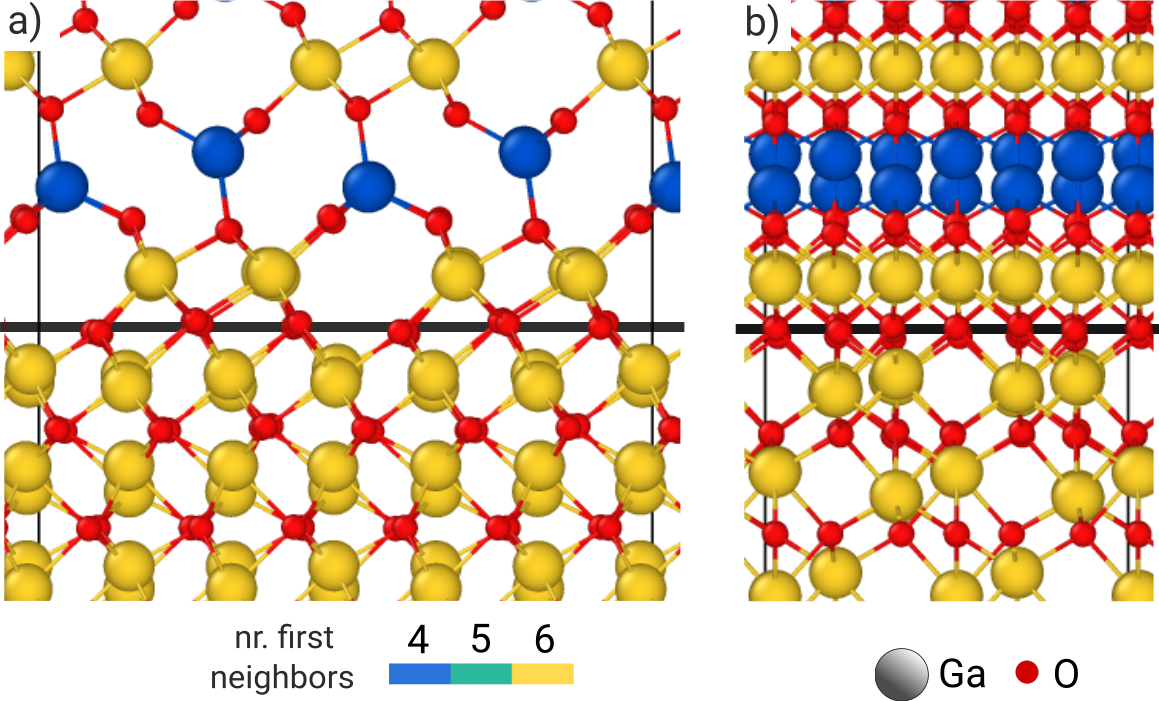}
 \caption{\label{fgr:Int b/a-Ga2O3_str} Front (a) and side (b) view of the $\mathrm{\beta-Ga_2O_3}$/$\mathrm{\alpha-Ga_2O_3}$ interface.  The $\mathrm{\beta-Ga_2O_3}$ interlayer is fully strained on $\mathrm{\alpha-Al_2O_3}$ substrate. The black line marks  the plane of O atoms shared by both film and substrate.}
\end{figure}

\begin{figure}[h]
\centering
  \includegraphics[height=5.2cm]{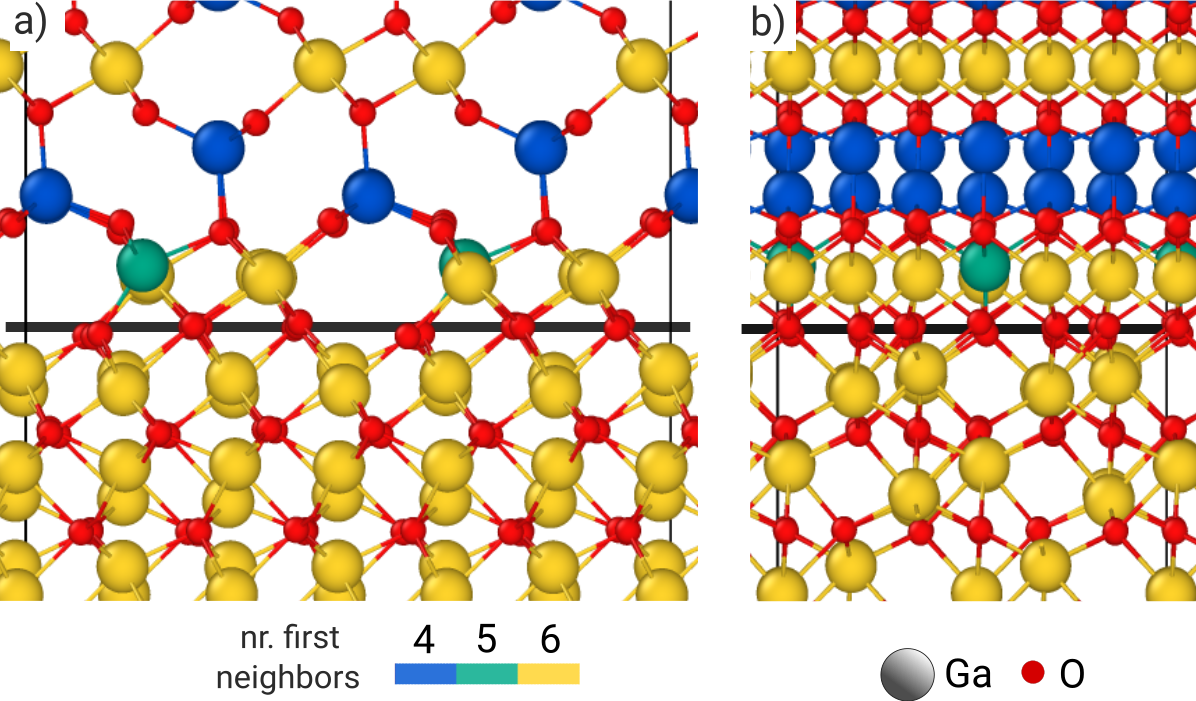}
 \caption{\label{fgr:Int b/a-Ga2O3} Front (a) and side (b) view of the $\mathrm{\beta-Ga_2O_3}$/$\mathrm{\alpha-Ga_2O_3}$ interface. The $\mathrm{\beta-Ga_2O_3}$ interlayer is fully relaxed. The black line marks  the plane of O atoms shared by both film and substrate.}
\end{figure}

\begin{figure}[h]
\centering
  \includegraphics[height=5.2cm]{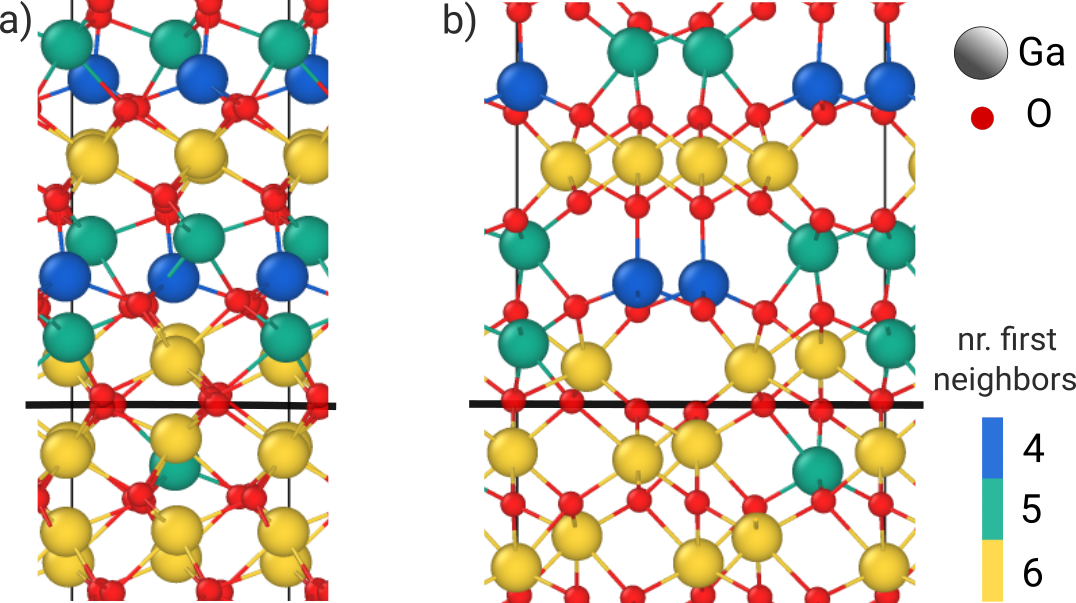}
 \caption{\label{fgr:Int k/a-Ga2O3}Front (a) and side (b) view of the $\mathrm{\kappa-Ga_2O_3}$/$\mathrm{\alpha-Al_2O_3}$ interface. The $\mathrm{\alpha-Ga_2O_3}$ interlayer is fully strained on $\mathrm{\alpha-Al_2O_3}$ substrate. The black line marks  the plane of O atoms shared by both film and substrate.}
\end{figure}

\begin{figure}[h]
\centering
  \includegraphics[height=5.2cm]{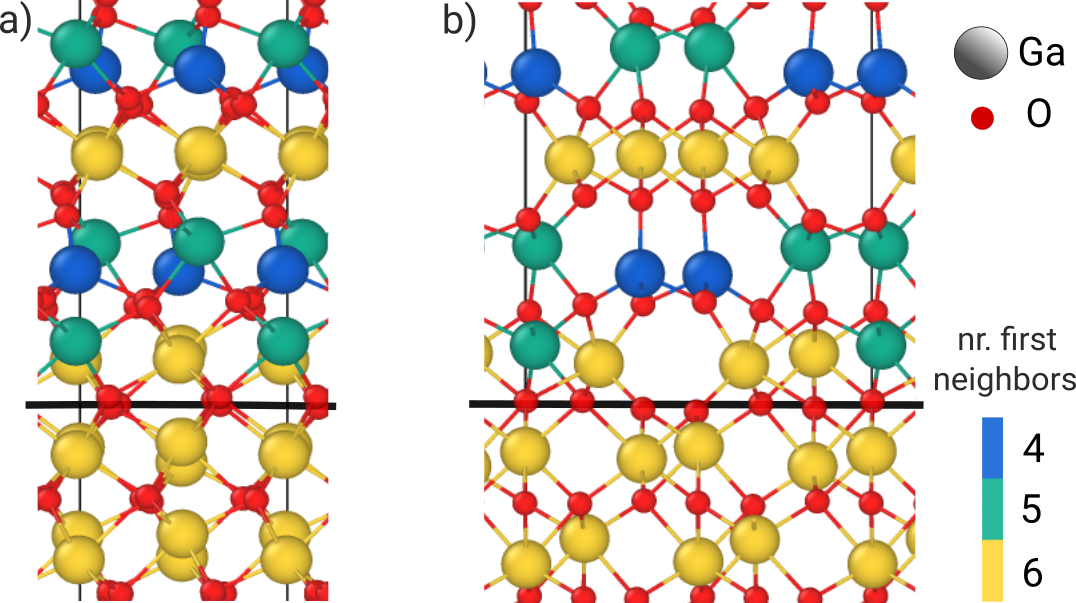}
 \caption{\label{fgr:Int k/a-Ga2O3_str}Front (a) and side (b) view of the $\mathrm{\kappa-Ga_2O_3}$/$\mathrm{\alpha-Al_2O_3}$ interface.  The $\mathrm{\alpha-Ga_2O_3}$ interlayer is fully relaxed. The black line marks  the plane of O atoms shared by both film and substrate.}
\end{figure}

\begin{figure}[h]
\centering
  \includegraphics[height=5.9cm]{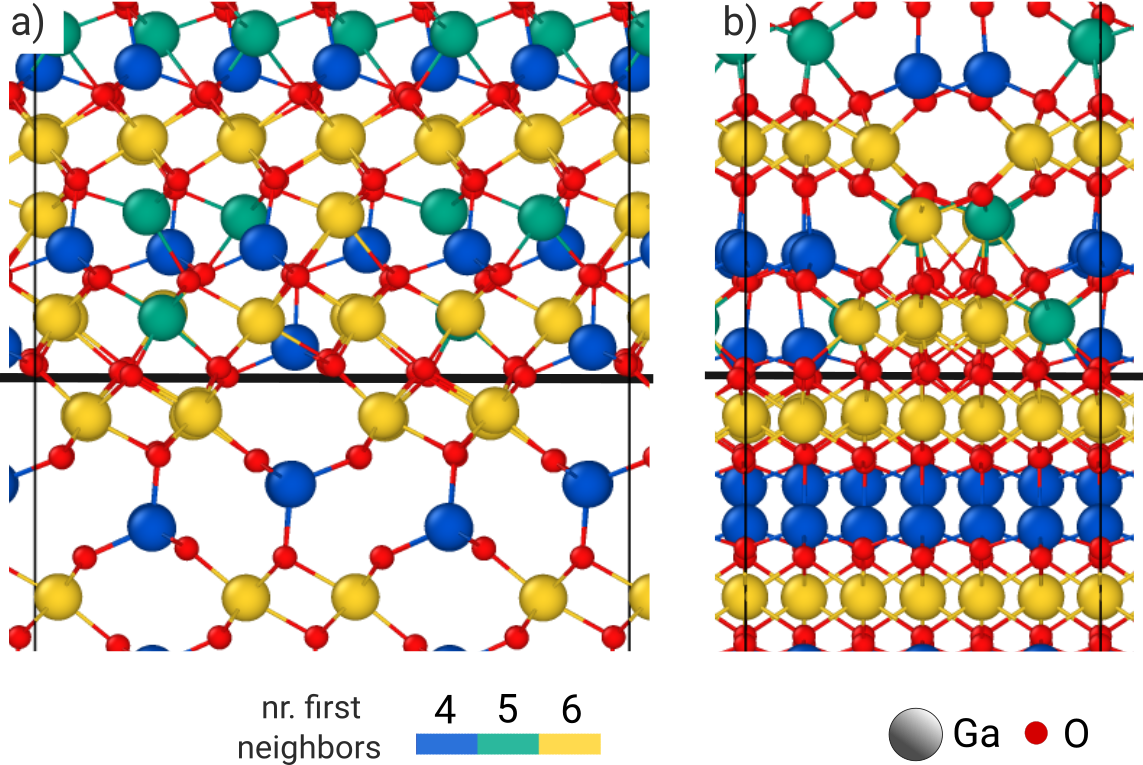}
  \caption{Front (a) and side (b) view of the $\mathrm{\kappa-Ga_2O_3}$/$\mathrm{\beta-Ga_2O_3}$ interface. The $\mathrm{\beta-Ga_2O_3}$ interlayer is fully relaxed. The black line marks  the plane of O atoms shared by both film and substrate.}
  \label{fgr:Int k/b-Ga2O3_rel}
\end{figure}
\clearpage
The files of the optimized structures reported in the main manuscript and in the Supplementary Information are collected in the archive supplementary\_information\_geometries.zip